\documentclass[preprint,5p,authoryear]{elsarticle}

\journal{Ecological Economics}

\usepackage{natbib}
\usepackage{graphicx}
\usepackage{bbold}    
\usepackage{mathrsfs} 

\newcommand{\ud}{\mathrm{d}}

\newcommand{\comment}[1]{}

\begin{document}

\begin{frontmatter}

\title{Notes from the Greenhouse World:\\
  A Study in Coevolution, Planetary Sustainability, and Community Structure}

\author[lw]{Lee Worden\corref{cor}}
\ead{wonder@riseup.net}

\cortext[cor]{Corresponding author}
\address[lw]{Environmental Science, Policy and Management,
  University of California, Berkeley, CA 94720-3114 USA}


\begin{abstract}

  This paper explores coevolution and governance of common goods using
  models of coevolving biospheres, in which adapting populations must
  collectively regulate their planet's climate or face extinction.
  The results support the Gaia hypothesis against challenges based on
  the tragedy of the commons: model creatures are often able to work
  together to maintain the common good (a suitable climate) without
  being undermined by ``free riders.''  A long-term dynamics appears
  in which communities that cannot sustain Gaian cooperation give way
  to communities that can.  This result provides an argument why a
  Gaia scenario should generally be observed, rather than a tragedy of
  the commons scenario.  Second, a close look at how communities fail
  reveals failures that do not fit the tragedy of the commons
  framework and are better described in terms of conflict between
  differently positioned parties, with power over different aspects of
  the system.  In the context of Norgaard's work, all these
  observations can be read as narratives of coevolution relevant to
  social communities as well as ecological ones, contrasting with
  pessimistic scenarios about common governance and supporting respect
  for traditional arrangements and restraint in intervention.

\end{abstract}


\end{frontmatter}


\section{Introduction}

James Lovelock and Lynn Margulis's Gaia hypothesis was aggressively
attacked almost immediately when it appeared in the 1970s.  The
hypothesis took a variety of forms ranging from ``atmospheric
homeostasis for and by the biosphere'' \citep{LovelockMargulis1974} to
``She [the earth] is now through us awake and aware of herself''
\citep{Lovelock1979}, and opponents tended to two lines of attack.
One was to dismiss Gaia talk as ``pseudoscientific mythmaking''
\citep{Postgate1988} because of poetic descriptions of the planet as a
living organism or a conscious, deity-like being.

The other was by means of the newly emerging sociobiological
vocabulary of ``self-interest'' and the ``common good.''  In the same
period of time, evolutionary biologists such as Richard Dawkins, E. O.
Wilson, R. L. Trivers, and G. C. Williams had launched a comprehensive
attack on the idea of the common good in evolution, exposing scenarios
in which selection on members of a collectivity acts against the
well-being of the collectivity those individuals belong to.  These
scientists successfully made the case that it is not sufficient to
explain or predict an evolved feature that serves a group of
organisms, or even a single organism, without explaining how selection
acting on the smaller parts --- organisms or their genes --- does not
destroy that feature.

Central narratives of this project, in addition to Dawkins's ``selfish
gene'' \citep{Dawkins1976}, are the prisoner's dilemma of game theory
\citep{RapoportChammah1965}, Garrett Hardin's tragedy of the commons
\citep{Hardin1968}, and the ``problem of collective action,'' made
formal by Mancur Olson \citep{Olson1965} and others.  These models
come originally from political science and economics, and language is
shared freely between evolution, politics and economics when they are
used.  Through this ``contact language'' \citep{Turner2006}, arguments
about these issues in biology tend to be indirectly about what is
possible in the social sphere as well.

The evolutionary attack on the Gaia hypothesis, introduced by W. Ford
Doolittle \citeyearpar{Doolittle1981} and taken up by Dawkins
\citeyearpar{Dawkins1982} and others, has come to be known as the
``problem of the population of one'' \citep{BarlowVolk1992}.  The
problem is that for the different species of the earth to collectively
regulate the atmosphere would be a kind of cooperation, or common
good, and so it must be explained why species' ``evolutionary
self-interest'' does not lead them to ``free ride'' and ``enjoy the
benefits'' of a well-regulated climate without contributing to it.
While this argument is closely related to Dawkins's discussion of
``selfish genes,'' it is best understood as a version of the tragedy
of the commons narrative structure, a connection which Dawkins makes
explicitly.  Populations of organisms stand in for animal herders, the
climate takes the place of the common pasture, and the ``temptation''
to accept the benefits of the common good without contributing is
predicted to lead to collapse.  This is called the ``population of
one'' problem in this context because while organisms are given
protection from destructive ``selfish genes'' by natural selection
that weeds out unhealthy organisms, natural selection cannot operate
on the whole biosphere because the biosphere does not reproduce and is
not part of a population of similar entities, leaving the problem of
how the biosphere could be protected from selfishness.

Attacks on the poetic-mythical language of Gaia advocates have
subsided, as key figures including Lovelock and Margulis have affirmed
their loyalty to science and cast their claims in the scientific
language of biogeochemistry and mathematical models.  The ``population
of one'' issue is still alive, and Lovelock has even conceded the
point \citep{Kerr1988}.

Lovelock, with Andrew Watson, introduced the Daisyworld family of
models to demonstrate that a planetary ecological community can act as
a global control system, keeping the climate within an optimal range
for living creatures \citep{WatsonLovelock1983}.  The impressive range
of Daisyworld models now in print does not resolve the ``population of
one'' question, because some Daisyworld communities lose their ability
to stabilize the climate when coevolution among the daisies modifies
their relationship to the climate, though others do not
\citep{Woodet2008}.

In this paper I will present a series of models that provide evidence
for ``sequential selection,'' a phenomenon that can produce planetary
ecological communities that stabilize themselves and the climate on
both ecological and evolutionary timescales.  Sequential selection is
a direct response to Doolittle and Dawkins' tragedy-of-the-commons-based 
challenges to the Gaia hypothesis.

A few recent modeling efforts
\citep{DowningZvirinsky1999,WilliamsLenton2007} have demonstrated
model communities that maintain environmental control on both
ecological and evolutionary timescales.  Williams and Lenton's
simulated biospheres achieve regulation via high-level selection
acting on an array of spatially distinct communities, a different
mechanism from the one seen here, and Downing and Zvirinsky's models,
while they demonstrate stable biospherical regulation without
high-level selection and are rich in fascinating behavior, are complex
and hard to analyze in detail, and the authors do not develop the
arguments that I present here.

In this paper I present results from a model framework that is
designed to have the richness of Downing and Zvirinsky's models,
together with as much as possible of the simplicity and mathematical
tractability of the Daisyworld models.  These models do not allow
selection among communities smaller than the whole biosphere.
Additionally, unlike previous models, in this framework the desired
climate is not specified by the modeler but negotiated by the model
organisms collectively.  These models explore when and how idealized
global ecological communities can achieve collective atmospheric
regulation.  Model simulation results indicate that the tragedy of the
commons scenario can happen, but suggest it is much less of a danger
than attacks on Gaia seem to imply.  Simple, uncontrived examples
arise in which such a tragedy is not an issue, and it becomes clear
how a global-scale self-organization process can produce a sequence of
persistent, self-regulating biospherical communities.

\section{Model Description}

In the models I present here --- the ``Greenhouse World'' family of
models --- the atmosphere is represented by a network of gaseous
compounds and processes that transform various compounds into each
other.  Transformations happen both spontaneously, by simple chemical
reactions, and as a byproduct of biological activity, like the
transformation of CO$_2$ into O$_2$ by plants and other
photosynthesizers.  Global temperature rises and falls depending on
the makeup of the atmosphere, and changing temperature helps or hurts
the creatures in the system.  Each population has an optimal
temperature: it can only survive temperatures within a few degrees of
its optimum, and produces offspring faster the closer the temperature
is to its optimum.  Simultaneously, the rise and fall of those
populations' reproductive rates changes the atmospheric balance, which
feeds back into the temperature.  Equilibrium requires whole-system
coordination of all these variables.  These models are not intended to
reflect actual climate dynamics realistically, only to provide a
somewhat tractable testbed for propositions about what is possible in
ecological communities.

Model equations are included in appendix~\ref{app:eco-eqns}.  The
technical details of these models are dealt with in detail elsewhere
\citetext{Worden and Levin, in preparation}.  Here I will present as
few equations as possible, and present an overview of conclusions from
the model experiments, explore further how exactly communities change,
and discuss the results in relation to coevolution and how
global-scale, sustainable ecological self-governance emerges and is
maintained.

On a slower timescale than the above ecological process, the model
species coevolve, responding together to their chemical and climatic
environment at the same time as they co-create it.  This is modeled by
a technique known by its proponents as adaptive dynamics
\citep{Metzet1996}: when the ecological dynamics are at rest,
introduce a single mutation --- a small variant population identical
to one of the established populations, except with a small, random
positive or negative number added to its optimal temperature --- and
continue the ecological dynamics.  Some variants thrive and old
populations decline to extinction, and long-term changes in the
community emerge.  (In this model context, I use the term `species'
somewhat arbitrarily for a lineage of populations, which is subject to
gradual evolutionary change as these variant populations arise and
replace their predecessors, and `community' for the collection of
coexisting species at a given time.)  In this model, variation is
limited to these species' optimal temperature, allowing them to adapt
incrementally to whatever temperature is currently found in the
atmosphere.

This process of continuous coevolutionary change eventually comes to
rest, and in extra-long-term experiments in qualitative change in
community structure, a new species is added to the network at this
point, with randomly assigned input and output resources and optimal
temperature, and the ecological and evolutionary process continues
from there.  As we will see, these simulation experiments yield a
significant conclusion --- the emergence of stably self-regulating
communities by means of sequential selection --- that directly
addresses the ``population of one'' question.

\section{The simplest community}

\comment{\it[Dawkins' question is posed in a model community comprised
  of only one population.  In this model system, the outcome is never
  tragic: an incentive structure emerges in which there are no free
  riders, as a population that changes its `strategy' so as to benefit
  more from the current climate also makes the climate more
  beneficial, not less.  This result underscores a conclusion I have
  reached in other work --- that collective action problems such as
  the tragedy of the commons scenario do not always arise and are only
  one of several different classes of incentive structures, and that
  incentive structures that induce ``free riding'' can sometimes be
  eliminated, averting tragedy.]}

\begin{figure} 
\centering
\includegraphics[height=1.5in,keepaspectratio]{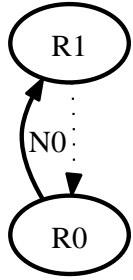}
\caption{Material flows in a simple model ecological community.  The
one population, N0, consumes resource R0 and produces R1 as a waste
product (solid arrow).  R1 degrades chemically to R0 (dotted arrow).
\label{fig:two-bubbles} }
\end{figure}

Figure~\ref{fig:two-bubbles} offers a very simple example of a model
community structure.  One species (labeled N0) consumes a gaseous
resource, R0 (its ``source resource''), and produces a different one,
R1, as waste.  Resource R1 spontaneously decomposes to R0, making a
sustainable cycle possible.  I assume R1 to be a greenhouse gas, such
that if the entire ecosystem were made of R1 the temperature would
rise to 100, and R0 to be a non-greenhouse gas, such that if
everything were R0 the temperature would drop to 0.  The shifting
balance between R0 and R1 places the temperature somewhere between
these two extremes.

\begin{figure}
\begin{center}
\resizebox{3in}{!}{
\includegraphics{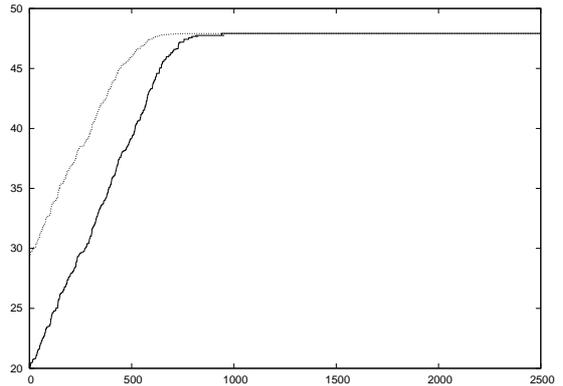} 
}
\end{center}
\caption{Evolution of optimal temperature in the model of
  figure~\ref{fig:two-bubbles}.  Time is on the horizontal axis,
  measured in number of mutations.  Upper (dotted) line is global
  equilibrium temperature, lower (solid) line is optimal temperature
  of population 0.}
\label{fig:two-bubbles-evolution}
\end{figure}

The development of this model community in evolutionary time is
plotted in figure~\ref{fig:two-bubbles-evolution}.  The species'
optimal temperature is initially 20, and it can survive temperatures
within 10 degrees of its optimum.  This initial system comes to an
equilibrium in which the atmosphere's temperature is nearly 30, which
is not optimal for the species, but is survivable.  As the species
evolves, its optimal temperature shifts upward, toward the actual
temperature, but each change in the species' population dynamics leads
to a shift in equilibrium such that the equilibrium temperature also
rises.  The difference between optimal and actual temperature stays
just under 10 degrees for a long time, but continually declines, and
ultimately the two temperatures meet at about 48 degrees.  When
species N0's optimal temperature coincides with the actual global
temperature, its evolution stops, and the system does not change any
further.

\subsection{Evolutionary dynamics in the simple case: no tragedy of
  the commons}

Now let us interpret the behavior of this model, first of all at the
beginning of its history.  This population requires a temperature
between 10 and 30, and temperature is determined by the atmospheric
balance that this population creates.  These organisms cannot conspire
(except in the etymological sense, meaning ``breathe together''); they
just dumbly produce offspring as quickly as the temperature allows
them to.  Without life, this planet's temperature would drop to 0.
The N0 creatures raise the temperature above 0 by exhaling a
greenhouse gas, R1.  As temperature rises above 20 and their
reproduction slows, we can imagine that~that slowing could either
speed up the rise in temperature further (a positive feedback) or
suppress it (a negative feedback).  Evidently it has a suppressing
effect, or the temperature would rise beyond 30 and kill the N0
population.  Similarly when temperature drops toward 20, the
population's response does not drive it further down, which would be a
destructive positive feedback.  Instead it raises the temperature,
stabilizing it at a survivable level.  In this sense self-regulation
prevails, and the birth rate and temperature come to balance each
other.  Should we view this as a counterexample to the tragedy of the
commons scenario, which predicts that global self-regulated harmony
should fail?

\begin{figure*}[!t]
\def\graphsize{0.9in}
\centering
\begin{minipage}[c]{\graphsize}
\begin{center}
\includegraphics[width=\graphsize,height=\graphsize,keepaspectratio]{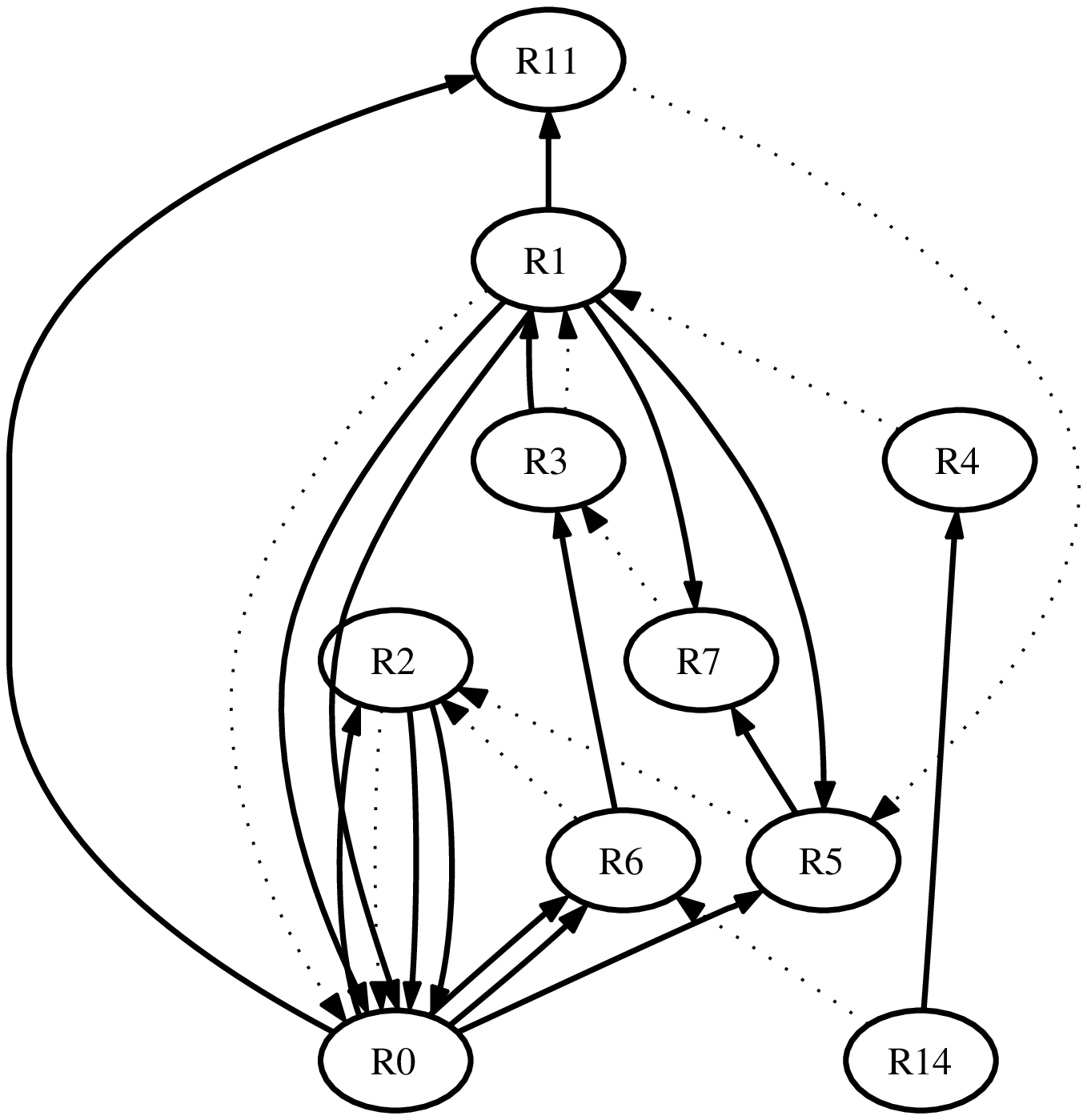}\\
\end{center}
\end{minipage}
\begin{minipage}[c]{\graphsize}
\begin{center}
\includegraphics[width=\graphsize,height=\graphsize,keepaspectratio]{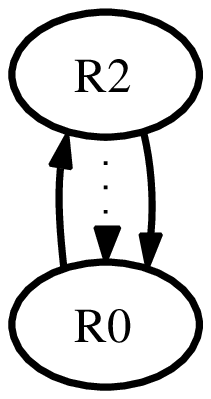}\\
\end{center}
\end{minipage}
\begin{minipage}[c]{\graphsize}
\begin{center}
\includegraphics[width=\graphsize,height=\graphsize,keepaspectratio]{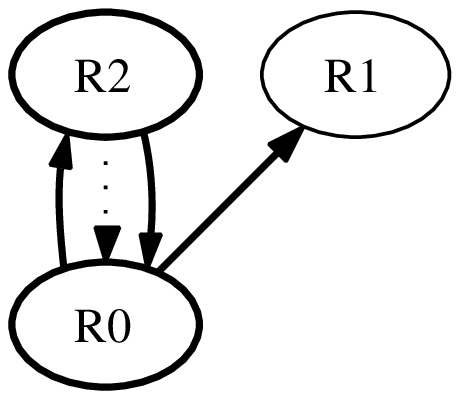}\\
\end{center}
\end{minipage}
\begin{minipage}[c]{\graphsize}
\begin{center}
\includegraphics[width=\graphsize,height=\graphsize,keepaspectratio]{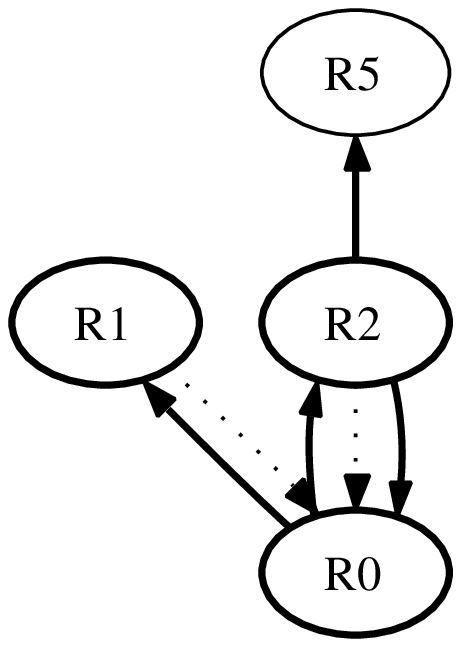}\\
\end{center}
\end{minipage}
\begin{minipage}[c]{\graphsize}
\begin{center}
\includegraphics[width=\graphsize,height=\graphsize,keepaspectratio]{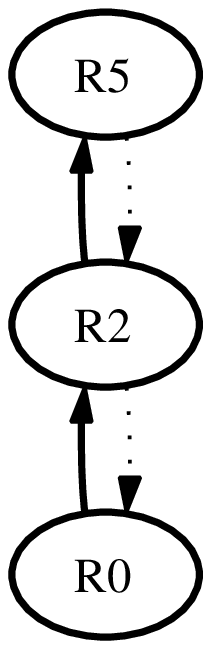}\\
\end{center}
\end{minipage}
\begin{minipage}[c]{\graphsize}
\begin{center}
\includegraphics[width=\graphsize,height=\graphsize,keepaspectratio]{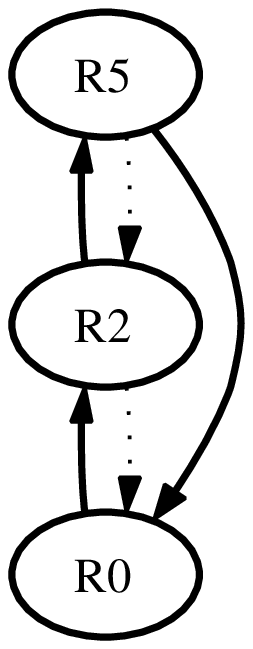}\\
\end{center}
\end{minipage}
\begin{minipage}[c]{\graphsize}
\begin{center}
\includegraphics[width=\graphsize,height=\graphsize,keepaspectratio]{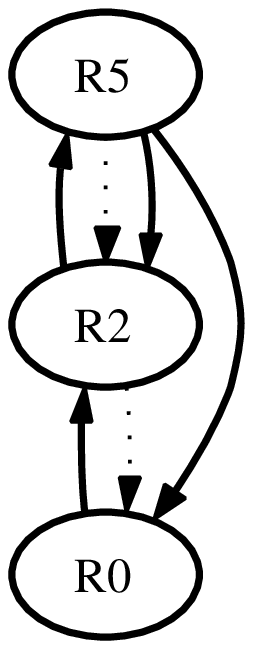}\\
\end{center}
\end{minipage}
\caption{Community structure changing as ecological need arises.  The
  first network pictured is randomly generated, and quickly collapses
  to the second structure shown.  Two new species are added in
  succession, the second of which brings about two extinctions.
  Further additions and collapses follow, and the community repeatedly
  regains climate equilibrium when it is lost.}
\label{fig:changing-communities}
\end{figure*}

That would be premature, because that prediction should be taken in
terms of evolutionary rather than ecological dynamics.  That is, given
a well-regulated biosphere, what is to prevent an antisocial
population from arising, which takes advantage of the good conditions
but does not help maintain them, making conditions worse for all
parties?  In this model, that would appear in the form of a variant
population that flourishes in the climate it finds, and as it grows in
number, either destabilizes the climate altogether, or leaves the
climate worse for all surviving organisms than it was beforehand.  The
difference between a species's optimal temperature and the actual
temperature is a convenient measure of species well-being in this
model framework.  Reproductive rate (i.e.\ fitness) is directly
determined by that difference.  Thus if a variant were either to drive
itself and the older type to extinction, or to increase the gap
between optimal and actual temperature --- which would lead to
``evolutionary suicide'' if repeated enough --- I would consider that
an example of a tragedy of the commons.

In this model, to the contrary, the temperature gap declines at every
step, until it reaches zero.  By the above logic, this is a
counterexample to the tragedy of the commons scenario.  It follows
that, at least in this model setting, the tragedy of the commons and
its related scenarios involving selfishness, free riders, and
defectors do not have predictive power.  ``Selfishness'' does not
necessarily undermine the sustainability of a common good.

In economics language, these tragic scenarios result from a perverse
incentive structure, which places the participants' short and long
term interests at odds with each other.  Incentive structures that
emerge in a given situation may or may not be perverse.  In each case
it is necessary to investigate whether perversity arises, and what
determines whether or when it arises.

\section{Complex communities: how sequential selection, driven by
  coevolution, can remove perverse incentives}

A randomly generated community of atmospheric compounds and species
that transform them is generally not viable as created.  Such an
assemblage tends to go through a period of climatic fluctuation and
one or more extinctions.  But an extinction changes the community to a
different, simpler, assemblage, that embodies a different circuit of
feedbacks.  That configuration may or may not be viable or stable, and
if not it will collapse to another community structure, which may or
may not do better.  In this way, the community will find its way to a
stable form by repeatedly revising itself, if it can find such an
arrangement before going to complete extinction.

In simulation experiments on such random communities, this search for
an ecologically stable subcommunity is usually successful, for a broad
range of choices of numerical parameters.  When such a community is
found, the process of coevolution may lead to further extinctions, or
the community may come to evolutionary stability without loss.  The
resulting community, if total extinction is averted, is stable on two
different time scales, both ecological and evolutionary.

Introduction of a qualitatively different species to an evolutionary
stable community may or may not disrupt the community, and if it does,
the above process of revision and re-stabilization repeats.
Figure~\ref{fig:changing-communities} presents the beginning of a
sequence of community structures generated by this cycle.  This
simulated community came to evolutionary stability 17 times, passing
through more than 25 network structures, before coming to total
extinction.

\subsection{Sequential selection weeds out dysfunctional communities}

This process in which inviable community structures give way to
viable ones by a sequence of restructuring events has been labeled
``sequential selection''
\citep{Lenton2004,Lentonet2004,BettsLenton2007}, by analogy to the
way natural selection removes poorly functioning organisms and
cultivates those that are able to stay alive and reproduce.  Like
Darwinian selection processes, sequential selection can produce systems
(communities, rather than organisms, in this case) that are not
sabotaged by the ``self-interest'' of the parts that make them up.
These ecological communities are stabilized by feedback processes in
which global atmospheric temperature is an essential component, making
the dynamic stability they exhibit inseparable from regulation of the
climate.

Thus, in the process of sequential selection, these communities
discard configurations in which climate regulation is undermined by
any of the species present, or otherwise ineffective, and organize
themselves into configurations where all species present contribute to
climate control, or least do not sabotage it.  In other words,
sequential selection can get rid of communities with conflictual
incentive structures, at least those that are bad enough to
destabilize the community, and select communities with generally
harmonious incentive structures.  This offers an answer to the
``population of one'' question of why the biosphere should not be
undermined by bad actors.  Sequential selection is very different from
Darwinian natural selection or other kinds of selection studied by
evolutionists (to be discussed more below), but like those processes,
it can counteract selection on smaller scales, producing a
larger-scale structure that is protected against ``selfishness'' of
its components.

\section{Coevolutionary dynamics in complex communities: cooperation,
  betrayal, and power}

\comment{\it[This section presents a model scenario with five coexisting
populations.  In this case (though not in the generic case), because
of the structure of the ecological network, only two of the five
populations determine the global temperature.  The other three
populations can adapt to the temperature but cannot affect it.  I
compare this situation to sociologists' theories of structural power,
in which agents' power is determined by their positioning in a network
of interactions.]}

\begin{figure*}
\centering
\def\wid{1.3in}
\begin{minipage}[c]{\wid}
\begin{center}
\includegraphics[width=\wid,height=\wid,keepaspectratio]{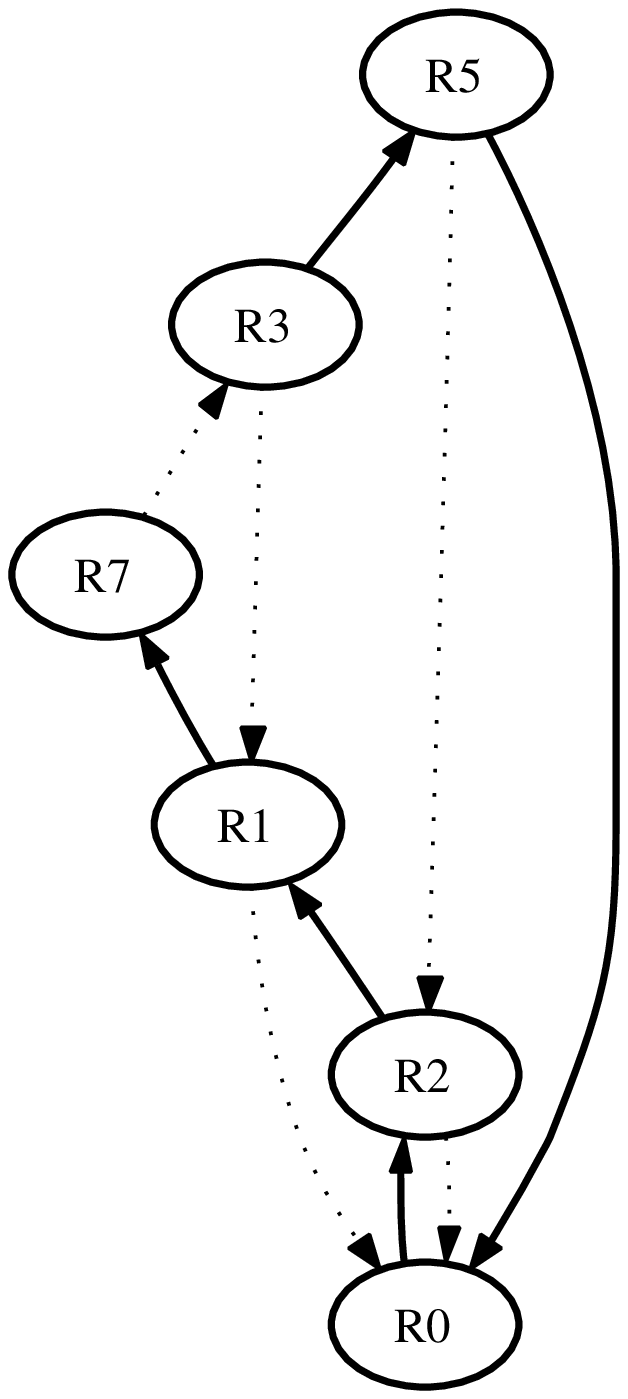}
\end{center}
\end{minipage}
\begin{minipage}[c]{\wid}
\begin{center}
\includegraphics[width=\wid,height=\wid,keepaspectratio]{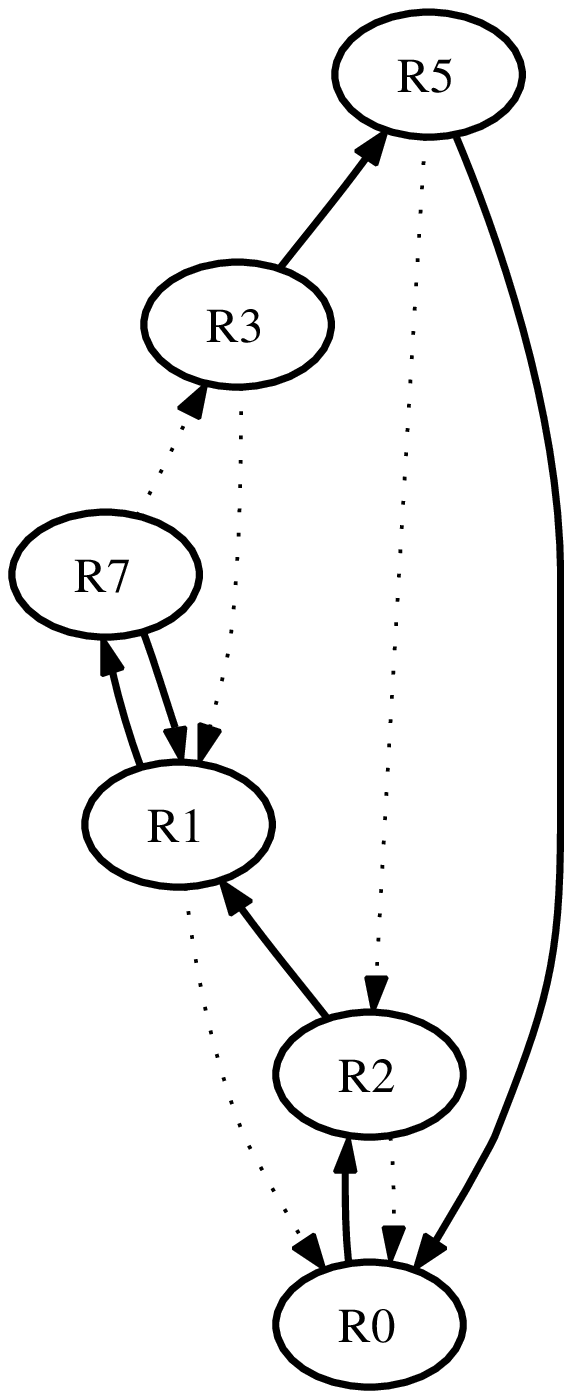}
\end{center}
\end{minipage}
\begin{minipage}[c]{\wid}
\begin{center}
\includegraphics[width=\wid,height=\wid,keepaspectratio]{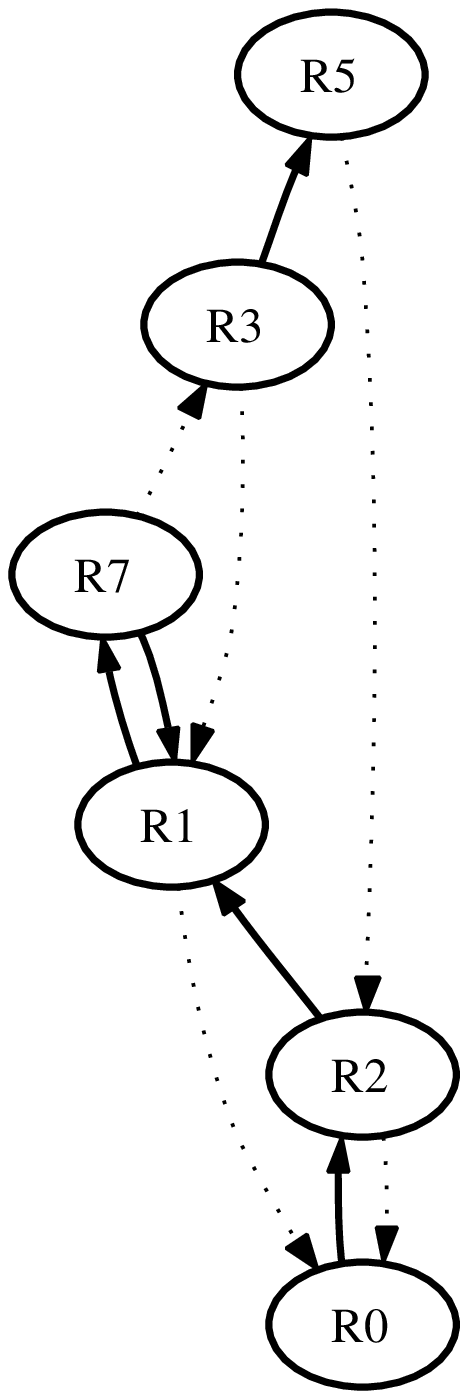}
\end{center}
\end{minipage}
\begin{minipage}[c]{\wid}
\begin{center}
\includegraphics[width=\wid,height=\wid,keepaspectratio]{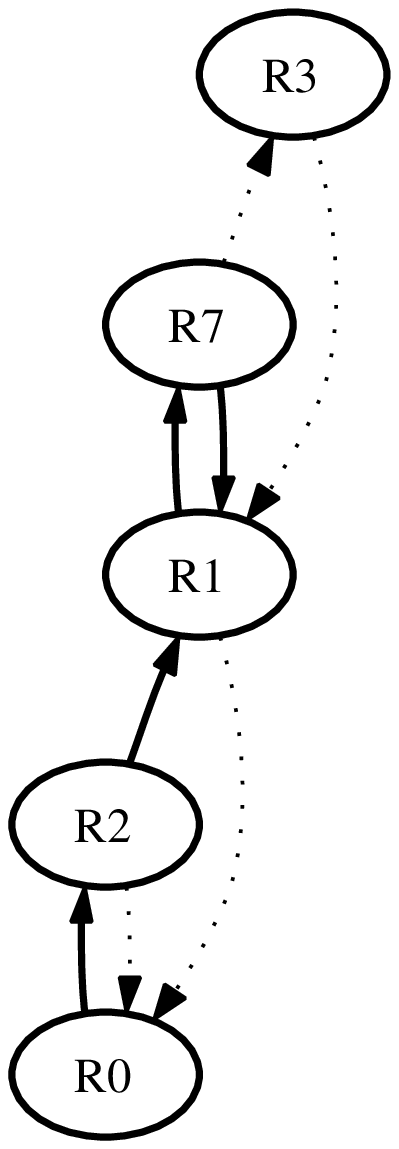}
\end{center}
\end{minipage}
\begin{minipage}[c]{\wid}
\begin{center}
\includegraphics[width=\wid,height=\wid,keepaspectratio]{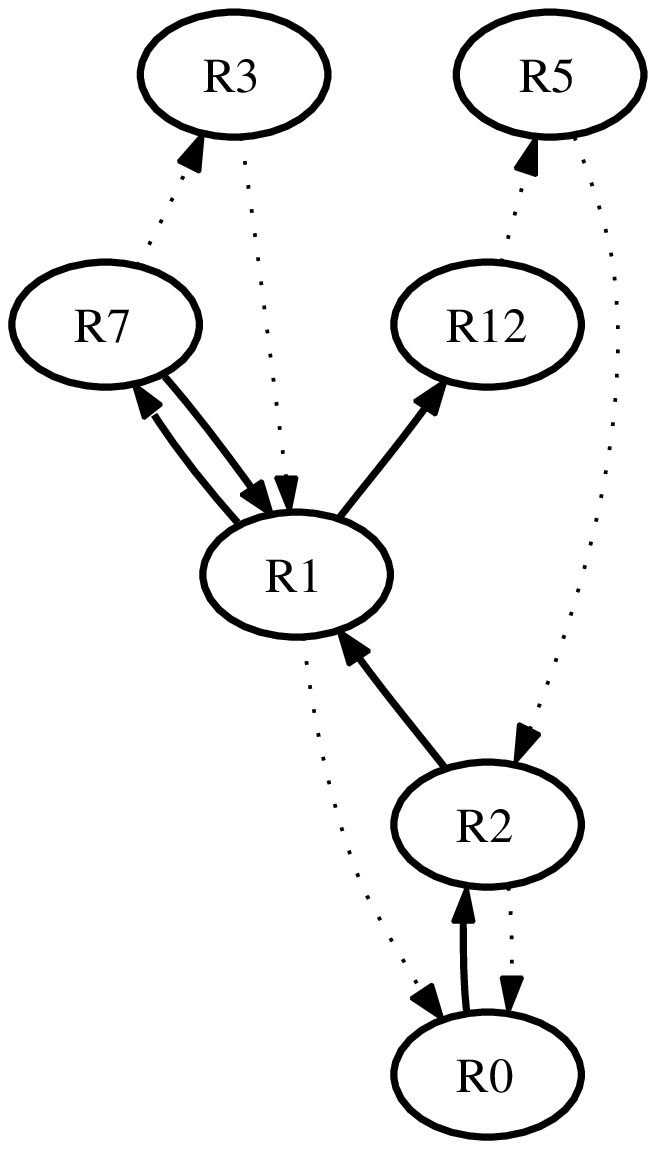}
\end{center}
\end{minipage}
\caption{Species introduction and subsequent changes in community
  structure.}
\label{fig:replacement}
\end{figure*}

These models tell a story of a sequence of stable communities,
punctuated by crises that flicker past as self-regulation is lost and
re-established.  Is it correct to describe this as periods of
cooperation, separated by temporary attacks of tragedy of the commons?
Let us look in some detail at how these communities fail.

First is an example where a new species is introduced and an old one
disappears.  The first three images in figure~\ref{fig:replacement}
show these changes in community network structure.

\comment{  The temperature veers away from its former equilibrium,
and then returns to almost exactly the same point.}

\comment{At the beginning of this crisis, the five species already present are
all adapted to the current temperature, about 58 degrees.  Resource R7
is significantly bigger than the others, because it is the only
resource without a consumer.  R7 is also more of a greenhouse gas than
most (if the entire biosphere were R7, the temperature would rise to
72 degrees), so it is keeping the temperature relatively hot, though
other resources in the system could do it if they were more abundant.
When the new species enters, consuming R7, it drains it out of the
air, causing a drop in temperature.  All the other species go into
decline in the cold, allowing all their source resources to
accumulate.  R3 in particular is an even stronger greenhouse gas than
R7, so this brings the temperature back up.}

When the new species enters this community, consuming R7 and producing
R1, resource R7 becomes far less available than before.  This reduces
the flow to the two resources and two species supplied by the decay of
R7.  The first of these two `downstream' species declines to a very
small population, and there is not enough left downstream of it to
sustain the second species in line, which goes extinct.

In dynamics terms, one would say that the old community's equilibrium
is made unstable by the introduction of the new variant, so that the
dynamics leaves that equilibrium point, finding its way to a different
stable attractor at which one of the old populations is reduced to
zero.  In ecological terms one might simply say that the new, invading
species displaces a resident species.  It is hard to see this as a
tragedy of the commons: is the new species more selfish than the old
one?  The new community is very similar to the old one, and both
stabilize at about the same temperature.  An essential aspect of the
tragedy of the commons is that the selfish party suffers along with
everyone else, which does not happen here.  We could say the new
species killed the old one, invoking images of warfare or crime, but
Hardin's narrative is a bad fit.

Later in this community's development, it acts out another scenario,
which we might call ``evolutionary murder,'' since one species is
killed by the evolutionary development of the others.  In this case,
the second of the two species downstream of R7 is also killed as a
result of the same invader, but this extinction happens well after the
introduction, as a result of coevolution of the species together, not
as a direct result of the introduction.  This unfortunate species,
like the one choking it off, needs to maintain its source resource at
a particular level in order to survive.  As the invader species
evolves, becoming better adapted to the climate, it uses its source
resource more efficiently and leaves even less of it in the air,
reducing the flow to the downstream species and eventually making it
unable to maintain enough source resource to survive.  (For the
mathematically inclined reader, because flows must balance at
equilibrium, the $R^*$ value for the upstream species equals the
downstream species's $R^*$ plus the downstream species's population
size [see appendices~\ref{app:rstar} and \ref{app:flows}], so that
when the two $R^*$s become equal the population vanishes.)

This extinction happens gradually, the dying species' population
becoming smaller with each evolutionary step of the upstream
population, and finally vanishing, while the overall community never
loses its dynamic stability or undergoes any discontinuous change.

The final collapse of this model planet's community results from a
catastrophic loss of biospherical self-regulation.  An invading
species takes over the R1 niche from a predecessor, displacing it and
another species that is fed by it (the two connecting R1 and R7).  The
gradual replacement of those two species by the new one brings with it
a drop in temperature, very slow and slight at first, but then
accelerating as the cooling hurts the remaining three populations and
causes them to release more of their source resources and produce less
of their products, which shifts the world much more drastically into a
cooler regime.  This is a runaway positive feedback that can not be
corrected as extinctions cascade.\footnote{For those who may be
  inspecting these model results in detail: this particular collapse
  seems to be a consequence of slight details.  It actually would not
  have happened if the separation of timescales were modeled
  perfectly, because in that case all the existing species would be
  perfectly adapted to the climate, and the invader would not be able
  to displace its competitor.  Instead, the existing consumer of R1 is
  almost perfectly adapted to the current climate, and the invader is
  just slightly better adapted.  However, other simulation
  realizations produce collapses that don't rely on tiny
  imperfections, since a newcomer can enter that uses an unused
  resource: there would still be collapses if this kind of imprecision
  were removed.}

This event does fit the tragedy of the commons mold.  The invading
species is responsive only to its own drive to reproduce, which has
the side effect of destroying the environmental conditions it and all
other life require for survival.  In simulations I have run, this
tends to happen after roughly 10 to 30 changes in community structure.
From earlier discussion about distinguishing between perverse and
cooperative incentive structures, one might think that each would
arise about half of the time, but that result indicates that perverse
structures, in the sense of those that destroy the community, are much
rarer than that.  Since each species here can only survive
temperatures within 10 degrees of its optimum, and temperatures range
from 0 to 100, this can be taken as evidence for a tendency for these
communities to survive by negotiating a stably harmonious order,
beyond random luck.

Also, it seems likely (though the complexity of the simulations has so
far precluded collecting sufficient data to test this hypothesis
directly) that as a community develops through addition and
subtraction of species, it has the potential to become immune to
further invasion.  When a community is fully evolutionarily
stabilized, a new species can only invade where there is a resource
with no consumer.  An evolutionarily stable community with an occupant
in every niche would be permanently safe from disruption.

\subsection{Ecological structural power}
\label{sec:power}

In the ``murder'' scenarios above, the only difference between
the murderer and its victims is their position in the ecological
network.  Network structure facilitates this kind of ``action at a
distance,'' and also creates ``keystone species'' like the primary
producer --- the species that metabolizes resource R0, without which
nothing else can survive because everything eventually decays to R0.  

These situations are reminiscent of sociologists' ``structural
power,'' in which certain players in a community have the ability to
determine outcomes and induce others to act in certain ways, by virtue
of their placement in a social network.  For example, if one is in the
only position bridging two disjoint subnetworks, one can become a
``broker'' who has a monopoly on information each group wants about
the other, or be especially creative by synthesizing ideas from the
two sides \citep[this is a structural hole, as in][]{Burt1992}.
Individuals who are well-connected ``network hubs'' may have
disproportionate influence over others.  The special positions found
in these Greenhouse World ecological networks are different from those
standard cases, and there may be an opportunity for fruitful
exploration of these kinds of structural effect in sociology or
economics, as well as in ecology.

\begin{figure}
\begin{center}
a.
\includegraphics[height=1.1in,keepaspectratio]{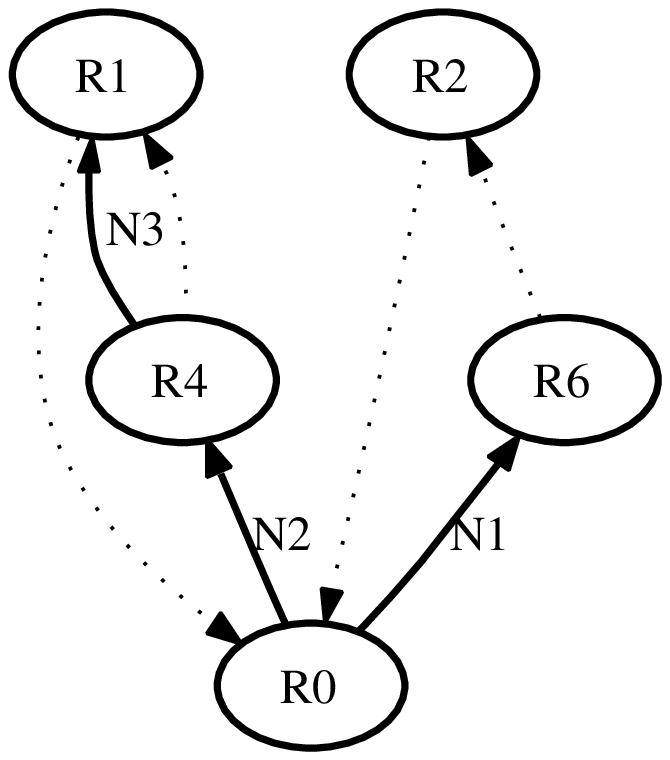}
\hspace{0.4in}
b.
\includegraphics[height=2in,keepaspectratio]{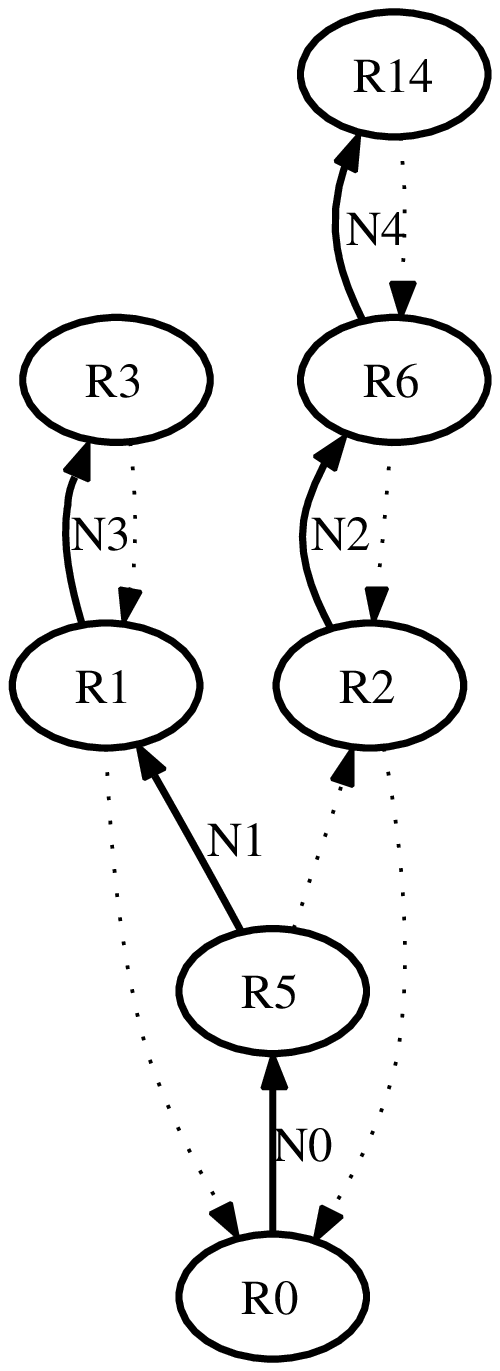}
\end{center}
\caption{a. In this network, because species N1 and N2 share resource
  R0 by controlling the temperature, species N3 has no impact on
  the equilibrium temperature. 
b. In this network, because species N1 and N2 are associated
  with the resources flowing between otherwise disconnected network
  components, they determine the climate, and the other species have
  no impact on it.}
\label{fig:control-graphs}
\end{figure}

There is also an exceptional condition that can arise in these
ecological models, which creates another kind of ``ecological
structural power.''  Every population in these models requires
temperature within 10 degrees of its preferred temperature.  When the
climate fluctuates, survival depends on the presence of a feedback
response to keep it within range.  Thus the ability of a population to
affect the temperature seems to be an important matter.  Generally,
all populations share control of the temperature, in the sense that
temperature shifts when any of them changes slightly.  However, there
are exceptional communities in which equilibrium temperature depends
only on two species out of a larger community.  I have seen this
happen in two ways.  These are illustrated in
figure~\ref{fig:control-graphs}.

In figure~\ref{fig:control-graphs}a, two species share a single source
resource.  This is generally not possible because of a rule of
competitive exclusion that emerges from the ecological dynamics (see
appendix~\ref{app:rstar}), but it can happen for a single pair of
species, if the temperature can be steered to the one exact value that
allows them to coexist, which is exactly halfway between those
species' optimal temperatures.

In the community of figure~\ref{fig:control-graphs}b, a similar
condition arises for different reasons.  Here two species bring the
temperature to a singular balancing point without sharing a source
resource.  This network is divided into two parts, connected only by
the flows from the source resources of the two species in question.
The flow from a species' source resource is directly related to the
difference between its optimal temperature and the actual temperature
(specifically, it is equal to that species' $R^*$, see
appendices~\ref{app:rstar} and \ref{app:flows}).  Since these two
flows must be equal at equilibrium, again, the equilibrium temperature
must fall exactly between these two species' temperature optima.

As a result of this concentration of ecological control of the
climate, as these communities develop coevolutionarily, the
equilibrium temperature always falls exactly between the two
controlling species' optimal temperatures.  As those species evolve,
the temperature changes accordingly, and other species must evolve
fast enough to keep up, or they will die, since they cannot affect the
temperature to keep it from going out of their range.\footnote{In
  practice this kind of extinction seems to be rare, since the two
  controlling species' optima, both evolving toward the actual
  temperatures, fall on either side of it, so the actual temperature
  tends not to change too drastically and other species are generally
  able to follow it successfully.  Without further investigation, it
  is not clear whether these cases are particularly important to
  understanding which communities live or die.}

\comment{\it[Can I get a case of being left behind by the temperature?]}

These special power positions are also different from the
sociologists' famous positions of structural power, and arise from
global rather than local processes, since these interactions are
mediated through the planetary temperature.

\comment{
\begin{figure}
\begin{center}
\end{center}
\caption{Actual and optimal temperatures vs. time in evolutionary
  murder scenario.}
\label{fig:evolutionary-murder}
\end{figure}

\begin{figure}
\begin{center}
\begin{minipage}[c]{1in}
\begin{center}
\end{center}
\end{minipage}
\begin{minipage}[c]{1in}
\begin{center}
\end{center}
\end{minipage}
\begin{minipage}[c]{1in}
\begin{center}
\end{center}
\end{minipage}
\end{center}
\caption{Community structures involved in evolutionary extinction.}
\label{fig:evolutionary-murder-graphs}
\end{figure}
}

\section{Sequential selection, multilevel selection, and social-cultural coevolution}

Sequential selection is very different from the selection processes
described by theorists of multilevel selection in evolution.  For
instance, group selection --- survival of groups that survive and
reproduce themselves better than others --- can overpower individual
selection in some cases when the two conflict, helping groups to
develop means to keep defectors or free riders in check, or simply not
have them \citep[see, for instance,][]{BoydRicherson2002}.  Similarly,
individual selection can work against gene-level selection, producing
organisms that have means to keep ``selfish genes'' in check or that
do not have such genes \citep[e.g.][]{WilsonWilson2007}.  While
sequential selection similarly interacts with lower-level processes,
producing planetary communities that are not undermined by ``selfish''
organisms, it is strictly speaking a theory of change and
stabilization in an isolated system, whereas multilevel selection
processes act in a population of similar, coexisting entities.

Strictly, then, if one were to try to apply multiple-level
evolutionary or coevolutionary narratives to the subject how local or
regional human communities change, such a discussion should probably
draw on ecosystem selection \citep{Swensonet2000a,Swensonet2000b} rather
than sequential selection, because of contact and competition between
different communities.  Regardless, the two have much in common, and
some of the conclusions of the present model may be of use in that
discussion, and particularly in relation to Norgaard's image of a
``coevolving cultural patchwork quilt.''

In a patchwork of communities, we can consider communities coevolving
with each other, and we can also consider each community being shaped
by coevolution among its constituent parts.  Sequential selection is
an aspect of the latter coevolution process.  It is emergent from the
dynamics of coevolution within a community in the same way that
Darwinian natural selection is emergent from ecological population
dynamics.

A long-persisting social/cultural/ecological community pattern can be
disrupted, or possibly even replaced, by individuals or social or
cultural patterns arriving from somewhere else.  Alternatively,
disruption can come from gradual or abrupt change within the
community.  In either case, the community must from time to time
either respond to perturbation, in accordance with its own internal
processes, so as to preserve its patterns of organization, or change
structurally.  It is reasonable to think that a period of change may
often be followed by arrival at a new pattern that has some level of
stability and persistence, and that qualities that correlate with such
persistent patterns are more likely to be observed at a given time
than qualities that correlate with brief periods of change.

\section{A Broader View of Ecological Governance}

In these ecological models, sequential selection produces communities
in which evolutionary incentives are aligned to stabilize coexistence,
unlike the tragedy of the commons scenarios proposed by Doolittle and
Dawkins.  When those stable arrangements do collapse, the dynamics of
the collapse also often do not fit the description of the tragedy of
the commons.

The fact that the tragedy of the commons narrative structure does not
accurately describe the events that shape these communities, which are
often better described in terms of conflict, together with the
possibility of an ecological analog of structural power, echoes
ecologist Peter Taylor's critique of the tragedy of the commons in
international politics \citep{Taylor1998,Taylor2005}: ``negotiations
and contestations among groups with different interests, wealth and
power---the messy stuff of most politics---are kept out of the
picture.  The `tragedy' thus naturalizes the liberalized economics of
structural adjustment and obscures the politics through which
structural adjustment is imposed and implemented in poor, indebted
countries'' \citep{Taylor1998}.  The ``negotiations and
contestations'' by which the climate is collectively navigated by
natural communities, when seen more clearly than the tragedy of the
commons narrative allows, may help open a door to an understanding of
``ecological politics'' that not only clarifies our ecological theory
but also helps to inform the crucial issues of how we may negotiate
our own issues of sustainability and coexistence.

\section{Conclusions}

\comment{\it[This section interprets the above and discusses what they
  suggest about human social dynamics and history, for instance how
  the Gaia scenario contradicts the tragedy of the commons scenario,
  and how the homeostat model resonates with Norgaard's coevolutionary
  narrative of history, in which social practices are adapted over
  time to local ecological conditions and vice versa.]}

The Daisyworld models are presented as a ``parable'' demonstrating
that Gaian regulation is a meaningful possibility
\citep{WatsonLovelock1983}.  The Greenhouse World models, like the
prisoner's dilemma and tragedy of the commons, can also be read as a
parable.  Here is one reading: common goods can in some cases be
maintained without tragedy; communities can turn out to be undermined
by selfishness far less often than pessimistic theories of altruism
and defection would suggest; and when a disruption arises in a
community, it can often be a matter of conflict between differently
positioned actors rather than a tragedy of the commons.  This
underscores the general points that while it is valid to ask whether
individual selfishness is a threat to any given common good, the
answer may well be that it is not, and that beyond the question of how
to overcome the problem of selfishness are the equally important
questions of when it is a problem and when it isn't, and how to change
the one case into the other.

These contentions are surely uncontroversial among many social
scientists and others who work to expose the ``messy stuff of
politics'' obscured by econ\-o\-mis\-tic narratives that deny
differences in power and opportunity.  Since evolutionary theory often
naturalizes narratives about selfishness, altruism, free riding, and
so on, and gives authority to proponents of individualistic,
ahistorical social theories, perhaps these results can be of use to
more institutionally or politically minded theorists, and in creating
space for exploring scenarios of collective governance that are not
supported by Hobbesian narratives of greed and competition.

This paper's theoretical results give support to the claims of Gaia
theorists: that planetary homeostasis by and for the biosphere makes
sense, and the planet can be expected to have some self-healing
mechanisms.  They also support some of the positions of skeptics, for
instance that the planet's harmony isn't invincible and can be
disrupted.  They comport with a view of the planet that avoids
extremes and is compatible with a contemporary understanding of
climate science: there are global feedbacks that produce the climate
we depend on, and they can be upset and give way to catastrophic
runaway positive feedbacks.  Both respect for the planet's
long-standing, slowly constructed traditions of ecological governance
and careful, responsible stewardship are called for.

The community structures predicted by sequential selection are not
just an accidental mush of miscellaneous elements.  They persist
because they are able to maintain themselves, in other words because
their overall structure is coherent, sustainable, and self-regulating.
That coherence may have taken a very long time to develop.  Disruptive
interventions can lead to large-scale crisis and to collapse into an
entirely different order.  This argument is the same whether
disruption comes from inside or outside the system.  It is an argument
for humility and respect for the wisdom of traditional practices and
arrangements.

These models additionally offer an argument against the use of tragedy
of the commons scenarios to justify ``development'' schemes such as
structural adjustment (there are many such arguments, but why not add
one more?).  Group selection, kin selection, reciprocal altruism,
social norms of reward or punishment, privatization or authoritarian
control need not be invoked to dispel the specter of common tragedy.
As in the curious little parables we have reviewed, a long-standing
social/cultural/ecological community of organisms and practices can,
by means of the self-organization process implicit in its history,
have become organized so that its constituents do not encounter
``temptation'' that lures them to antisocial acts.  Actually,
disruption of an established order can introduce a destructive
incentive structure, with the expected tragic consequences.

\comment{\it To do: How sequential selection is driven by coevolution;
  Norgaard connections}

\comment{\it[Sequential selection shouldn't be looked for casually in human
communities, because they aren't isolated.  Social researchers
interested in evolutionary narratives should be conscious of levels of
selection.]}  

\comment{\it[A blow to the tragedy of the commons.  Popular claims about
selfishness and individuality are exaggerated.  Higher level patterns
can be selected.  A bit about community-level selection.]}

\comment{\it[@@ Mind and nature: selection and learning.  Reductionism
  hasn't killed cybernetic holism.]}

\comment{\it[How higher-level order in communities argues for preservation and
local knowledge, against World Bank-style standardization.  The
population of one argument in the context of individualistic
free-market politics.]}


\section{Acknowledgements}

This research and writing have been conducted during the course of an
S.~V. Ciriacy-Wantrup Postdoctoral Fellowship at UC Berkeley,
supervised by Richard Norgaard in the Energy and Resources Group, and
Ignacio Chapela in Environmental Science, Policy and Management.  I am
very grateful to all those mentioned above, and also to Simon Levin
and my Berkeley study group, for many useful and enjoyable
discussions.

\bibliographystyle{elsarticle-harv}
\bibliography{master-noplus}

\newpage
\appendix
\renewcommand\thesection{A}
\section*{Appendices}

\comment{\it[Note: these appendix sections could also be sidebars, or
  floating boxes]}

\subsection{Model equations}
\label{app:eco-eqns}

Population sizes are represented by $N_i$, for $i=1$ to $n_p$;
resources (atmospheric compounds) are $R_j$, for $j=1$ to $n_r$; and
global temperature is $T$.  Each population has an optimal temperature
$\tau_i$.  Intrinsic population growth rate,
dependent on temperature, is
\[
 r_i(T) = R_{s(i)} u(T,\tau_i) =
   \left\{\begin{array}{lll} R_{s(i)} r_0
       \left(1-\frac{\left({T-\tau_i}\right)^2}{\sigma^2}\right)
       & \quad & \mbox{if positive,} \\
     0 & & \mbox{otherwise.}
   \end{array}\right.
\]
Here $s(i)$ indexes the source resource of population $i$, meaning the
one it consumes, $r_0=50$ is a rate constant for population growth,
and $\sigma=10$ is the maximum temperature deviation the population
can tolerate. [In figure~\ref{fig:changing-communities} $\sigma$ is 16
rather than 10.]

Each resource is assigned a `greenhouse constant,' or heating
coefficient, $H_j$.  Additionally, $p(i)$ indexes the resource
produced by population $i$ as waste, $\gamma=1/2$ is the populations'
efficiency of uptake of their source resources, $m=1$ is the
populations' mortality rate, $\rho=1$ is rate of spontaneous chemical
reduction, $\mathit{red}(j)$ is what resource $j$ reduces to,
$\Lambda=1000$ is the rate of temperature change in response to the
greenhouse effect, and $M$ is the total mass of populations and
resources (which is constant over time).

The model ecological dynamics are then
\begin{eqnarray*}
  \frac{\ud N_i}{\ud t} & = &
             \gamma\ r_i(T)\ N_i - m\ N_i \\
  \frac{\ud R_{\!j}}{\ud t} & = & 
     \!\!\!\!\! \sum_{\{i| p(i)=j\}}
               {\!\!\!\!\left(1-\gamma\right)\ r_i(T)\ N_i}
                              \\
            & & \quad \mbox{} + \!\!\!\! \sum_{\{i| s(i)=j\}}
           {\!\!\!\!\!\left(m - r_i(T)\right)\ N_i}
                              \\
            & & \quad \mbox{} +\!\!\!\!\!\! \sum_{\{j'| \mathit{red}(j')=j\}}
		{\!\!\!\!\!\!\!\rho\ R_{j'}}\ -\ \rho\ R_{\!j} \\
  {\ud T\over{\ud t}} & = &
    \Lambda \left( \sum_j \frac{H_{\!j}\ R_{\!j}}{M}
	                          - T\right).
\end{eqnarray*}

\subsection{The $R^*$ result}
\label{app:rstar}

Population sizes and resource concentrations at equilibrium, in
general, are hard to analyze in this system without solving for all
the system variables together.  However, there is a simple result for
the source resource of a population.  Equilibrium conditions are found
by equating the model dynamics equations (appendix~\ref{app:eco-eqns})
to zero.  In the case of the population dynamics equation, that gives
\[ \gamma\ r_i(T)\ N_i - m\ N_i = 0, \]
which is solved either when $N_i=0$, or when
\[ R_{s(i)} = \frac{m}{\gamma\ u(T,\tau_i)}. \]
This means that for any population that is not extinct at equilibrium,
its source resource must equilibrate at the above level, which we can
call $R^*(T,\tau_i)$.

It follows that if two populations have the same source resource but
different $\tau$, since the source resource cannot equilibrate at two
different levels, generally one of the two must go extinct.  This
result will be familiar to students of Tilman's work on resource
competition in ecology \citep{Tilman1982}.

\begin{figure}
\begin{center}
\resizebox{3in}{!}{
\includegraphics{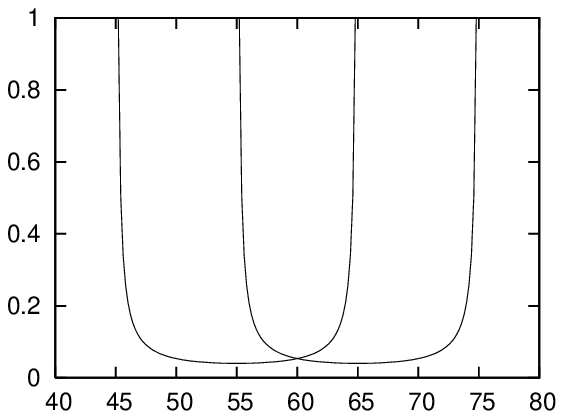} }
\end{center}
\caption{For two species with different $\tau$, $R^*$ is equal at only
  one temperature.}
\label{fig:rstar-crossing}
\end{figure}

In this system, though, there is an exception to the standard $R^*$
result, since $R^*$ is a function of temperature, and population
dynamics can change the temperature: if the temperature can be brought
to exactly the value where two populations' $R^*$ curves cross
(figure~\ref{fig:rstar-crossing}), those two populations can coexist
at equilibrium.  This is a real possibility, as discussed in
section~\ref{sec:power}.  The crossing point is halfway between the
two populations' optimal temperatures.  For obvious reasons, when
optimal temperatures are not specially chosen, this can only happen
for one pair of populations out of the entire community at one time.

\begin{figure}
\begin{center}
\resizebox{3in}{!}{
\includegraphics{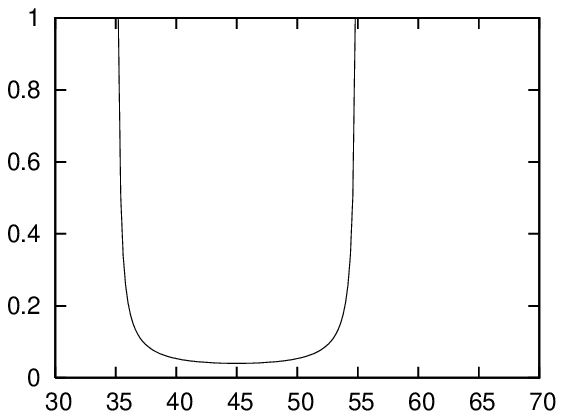} }
\end{center}
\caption{$R^*(T,\tau)$ as a function of $T$, for a population whose
  optimal temperature $\tau$ is 45.}
\label{fig:rstar}
\end{figure}

\subsection{Balancing flows}
\label{app:flows}

If we examine the resource dynamics equations at equilibrium, as we
did with the population dynamics equations to get the $R^*$ result, we
learn how to associate the population and resource variables with
flows across the network.  The $R^*$ calculation tells us that at
equilibrium $r_i(T)=m/\gamma$.  Using that fact, the resource
equation at equilibrium reduces to
\def\back{\!\!\!\!}
\[
0 = \back\sum_{\{i| p(i)=j\}}\back N_i\ - \back\sum_{\{i|s(i)=j\}}\back N_i\ 
   + \back\!\!\!\sum_{\{j'| \mathit{red}(j')=j\}}\back\!\!\! R_{j'}\ -\ R_{\!j}.
\]
The equilibrium flows across the arrows in a network are equal to the
population size, for a (solid) arrow representing flow due to a
population, or abundance of the decaying resource, for a (dotted)
arrow representing decay of a resource.  This is a balance equation,
expressing that the total flow into a node of the network must equal
the total flow out of it.

\end{document}